\documentstyle[preprint,revtex]{aps}
\begin{document}
\draft
\begin{title}
Hole-hole correlations in the $U=\infty $ limit of
the Hubbard model \\
and the stability of the Nagaoka state
\end{title}
\author{M.W. Long}
\begin{instit}
School of Physics \\
University of Birmingham \\
Edgbaston, Birmingham B15 2TT, United Kingdom
\end{instit}
\moreauthors{X. Zotos}
\begin{instit}
Institut Romand de Recherche Num\'erique en Physique des Materiaux
(IRRMA)\\
PHB-Ecublens, CH-1015 Lausanne, Switzerland
\end{instit}
\bigskip\bigskip
\begin{abstract}
We use exact diagonalisation in order to study the infinite - $U$ limit of the
two dimensional Hubbard model.
As well as looking at single-particle correlations, such as
$n_{{\bf k}\sigma }=\langle c^\dagger _{{\bf k}\sigma }c_{{\bf k}\sigma }
\rangle $, we also study {\it N-particle correlation functions} which
compare the relative positions of {\it all} the particles in different models.
In particular we study 16 and 18-site clusters and compare the charge
correlations in the Hubbard model with those of spinless fermions and
hard-core bosons. We find that although low densities of
holes favour a `locally-ferromagnetic' fermionic description, the
correlations at larger densities resemble those of
pure hard-core bosons surprisingly well .
\end{abstract}


\newpage

In the following we will study the hole-hole correlations in the t-model
given by the $J \rightarrow 0$ limit case of the Hamiltonian:

\begin{equation}
H=-t\sum_{\langle ij\rangle \sigma }(\tilde c_{i\sigma }^\dagger
\tilde c_{j\sigma }+ h.c) +
J \sum_{\langle ij\rangle }(\vec S_i \cdot \vec S_j -n_i n_j/4)
\end{equation}

\noindent
where $\tilde c_{i\sigma }(\tilde c^{\dagger}_{i\sigma })$
are annihilation (creation) operators of a fermion on site i with spin $\sigma$
and the tilde implies the restriction to single occupancy.
The sum is over all bonds $\langle ij\rangle $ of a two dimensional lattice
with periodic boundary conditions.

This Hamiltonian is the generic model for a doped antiferromagnet and
also a true Hubbard model in the limit $U=\infty $.  We have chosen
the case with no spin fluctuations because a finite $J$ induces
attraction between
the holes forming pairs (for larger values of $J$ even phase separation) and we
want to study the simple effect of the existence of a spin background on the
motion of the holes.  Although it is by no means well established, it seems
reasonable that we can interpret the t-model in terms of a competition between
two effects:  Firstly, we have an interaction which promotes ferromagnetism
locally around an isolated hole, an effect resulting from the physics
inherent in Nagaoka's Theorem \cite{nag},
and secondly we have an interaction which
promotes low-spin, coming from the desire for a low-density of particles to
have a paramagnetic ground state (see Kanamori \cite{kan}).
We will refer to the
saturated ferromagnetic state indicated by the first phenomenon as the
`Nagaoka state'.

We consider a lattice with N
sites, away from half-filling with $N_c$ fermions and $N-N_c\equiv N_h$ empty
sites (or holes).
In a previous work \cite{lx} we presented a
variational argument indicating that the
Nagaoka state should be unstable and the holes should have correlations and
energy close to that of hard core bosons (h.c.b.).
The argument is as follows; the ground state wavefunction
of the combined hole-spin system can always be factorized into the product of
a charge part times a spin part.

\begin{equation}
|\Psi>=\sum_{\{ c \}} \alpha_{c} |c> ( \sum_{\{ s_c\} } \beta_{s_c} |s_c> )
\end{equation}

\noindent
where the first sum is over all charge configurations $|c>$ and the second sum
is over all spin configurations $|s_c>$ for the given hole configuration $|c>$.
Further we have chosen the factors ${\alpha_c}$ positive and the coefficients
${\beta_{s_c}}$ such that:

\begin{equation}
\sum_{\{s_c\}} |\beta_{s_c}|^2=1
\end{equation}
\noindent
We can then define a unit vector in spin space ${\vec \Sigma}_c=
\sum_{\{ s_c\} } \beta_{s_c} |s_c>$. One way to implement this description is
in terms of a `slave-fermion' representation: $c^\dagger _{i\sigma }=f^\dagger
_ib^\dagger _{i\sigma }$ where $f^\dagger _i$ is a pure fermion operator which
carries the statistics and $b^\dagger _{i\sigma }$ is a pure boson which
carries the spin.  In these terms $|c>=\prod _{i\epsilon I}f^\dagger _i|0>$ and
$|s_c>=\prod _{i\epsilon I}b^\dagger _{i\sigma _i}|0>$, corresponding to the
state with electrons on the sites contained in $I$ with spins $\sigma _i$.
In this representation the ground state energy $E$ can then be written as:

\begin{equation}
E=<\Psi|{\hat T}|\Psi>=\sum_{\{c',c\}} \alpha_{c'}\alpha_c
<c'|{\hat T}|c> {\vec \Sigma}_{c'}\cdot {\vec \Sigma}_{c}
\end{equation}

In this expression the charge configurations $|c> , |c'>$ are related by the
hop of a single particle between neighboring sites with amplitude
$<c'|{\hat T}|c>=\pm t$ because
of the fermionic character of the particles. The product in spin space
${\vec \Sigma}_{c'}\cdot {\vec \Sigma}_{c}$ can take values between
$\pm 1$. In the case of the completely ferromagnetic state the spin
configurations ${\vec \Sigma}_c$
are `parallel', so the product is always +1 and the energy of the system is
that of a Fermi sea filled with same-spin fermions. In principle however the
system could minimize its energy by using the spin degrees of freedom which
we can consider as variational parameters. In the ideal case whenever the
sign of the hopping is +t the `spin wavefunctions' should be `antiparallel'
thus cancelling the fermionic sign exactly. In this case then the holes would
have exactly the same energy and charge correlations
(expressed by the $\alpha_c$'s) as free hard core bosons.
We should note that as the spin fluctuation term is zero there is no energy
associated with spin flips.
Actually as we will see this `fermion' to `bosons' transformation works
perfectly for a system of a single square plaquette because the spin
configuration in this case corresponds to a singlet.

To decide about the behavior of holes in the presence of the spin background
we studied, besides the energy, the hole-hole correlations. In particular we
want to compare the hole-hole correlations in the t-model with the
correlations of particles in different reference models, such as hard core
bosons
or spinless fermions. As a measure of the resemblance we defined an `overlap'
between the hole-hole correlations in the t-model and the reference model as:

\begin{equation}
(t|r)=\sum_{i_1<i_2<,...,<i_{N_h}}
\sqrt{ <\Psi_t|n_{i_1} n_{i_2}...n_{i_{N_h}}|\Psi_t> } \cdot
\sqrt{ <\Psi_r|n_{i_1} n_{i_2}...n_{i_{N_h}}|\Psi_r> }
\end{equation}

\noindent
where $n_i=1-f^{\dagger}_i f_i=1-n^c_i$ is the hole number operator
at site i in the t-model or
the particle number in the reference model and $N_h$ is the number of holes.
$|\Psi_t>$ is the ground
state wavefunction of the t-model and $|\Psi_r>$ of the reference model.
It is easy to see that this overlap  $\le +1$, taking the
value +1 when {\it all} the charge correlations are identical.
We should also note that the factors $\alpha_h\equiv \alpha _c=
\sqrt{ <\Psi_t|n_{i_1} n_{i_2}...n_{i_{N_h}}|\Psi_t>}$ as defined above,
since $<\Psi_t|n_{i_1} n_{i_2}...n_{i_{N_h}}|\Psi_t>=
<\Psi_t|n^c_{j_1} n^c_{j_2}...n^c_{j_{N_c}}|\Psi_t>$, where $i_n$ are the
sites of the holes and $j_n$ are the complementary sites of the particles.
This overlap is a rather global comparison of the
hole-hole correlations between the t-model and the reference model and is not
extremely sensitive as the differences enter as quadratic effects; viz
$\sum_{\{c\}}
\tilde \alpha _c\cdot \alpha _c=1-{1\over 2}\sum_{\{c\}}
(\tilde \alpha _c-\alpha _c)^2$.

We will first present the results for different size systems, from 4 to 18
sites and different hole densities. They were obtained by exact
diagonalization of the Hamiltonian using the Lanczos technique.  For the
present N-particle correlation functions it is by no means clear whether or
not Quantum Monte Carlo techniques can evaluate these quantities.
As we discussed above, for the system with 4 sites and 2 holes the
ground state is a singlet ($S=0$) and the energy is $E=-2\sqrt{2}$,
the same as that for hard core bosons.
The simple explanation of this fact is the observation
that for two fermionic holes exchanging in the presence of a singlet, they
exchange
as hard core bosons, the fermion minus sign cancelled by the rotation of the
singlet wavefunction. Another point of view is that the singlet
spin wavefunction corresponds to a rotation of the boundary conditions which
shifts the allowed $k$ vectors from the set $\{0,\pm\pi/2,\pi\}$ to
$\{\pm\pi/4,\pm 3\pi/4\}$.

The next system where this transformation from fermion to boson correlations
is exact is the 8 site system. In this case one of the ground
states is a singlet ($S=0$) with
energy $E=-2 \sqrt{2} \sqrt{3}=-4.8990$.
For comparison the energy for 2 hard core bosons in this lattice is
$E_{h.c.b}=-4 \sqrt{3}=
-6.9282$ and for 2 spinless fermions $E_f=-4$.
What is interesting in this system is that the hole-hole correlations
are identical to those of two hard core bosons,
overlap (t$\mid$h.c.b)=+1, although the energy is much higher (rather
closer to that of free spinless fermions -the Nagaoka state-).
This observation is, as we show below, true for almost all systems we studied.

In Tables \ref{table1},\ref{table2} we present the
energy E and `overlap' (t$\mid$h.c.b) of the
hole-hole correlation functions to those of hard core bosons as a function
of $S^z$, the number of spin flips from the totally ferromagnetic state
-the Nagaoka state -, for different number of holes $N_h$.
As we have mentioned above our criterion of the `overlap' of correlation
functions is not sensitive to the details, so on the right hand side of these
tables we also give the hole-hole pair correlation function
$g(\vec r)_t=<n_0 n_r>$ for different number of holes in the ground state as
well as those of hard core bosons $g(\vec r)_{h.c.b}=<n_0 n_r>$in the ground
state.

We should note that we present results for number of holes where the ground
state is a singlet; in these cases the momentum of the ground
state is $\vec k=(0,0)$ and therefore the wavefunction isotropic,
the same as that of the hard core boson system.
However for $S^z \ne 0$ the lowest energy state can be degenerate with
$\vec k \ne (0,0)$ corresponding to an anisotropic wavefunction rendering the
comparison to hard core bosons more tentative.
In general the behavior of the ground state spin for different
numbers of holes is rather erratic (for further results see ref.
\cite{ry,fe}),
probably due to competition and finite size effects, as we will discuss below.
Interesting is also the case of 16 sites with 5 holes, corresponding
to the case of a `closed shell',  where the
energy is not a monotonic function of $S^z$
initially increasing with the first spin flip and then decreasing,
with the ground state at $S=1/2$.

Comparison with free spinless fermions gives an overlap for N=16 sites:
(t$\mid$s.f)=0.9211, 0.8899, 0.8247 for $N_h$=2, 4, 6 respectively,
while for N=18
sites: (t$\mid$s.f)=0.9373, 0.9129 for $N_h$=2, 4. Finally comparing with a
Gutzwiller projected wavefunction for 10 fermions in 16 sites
(a nondegenerate, $\vec k=(0,0)$ ground state) we obtain
an overlap of 0.9695; so for both a spinless fermion and a Gutzwiller
wavefunction we find an overlap significantly smaller than that of
hard core bosons.

In an attempt to get an insight into the spin part of the wavefunctions
we also calculated the Fermi distribution function $n_{\sigma}(k)=
<\tilde c^{\dagger}_{k\sigma}
\tilde c_{k\sigma}>$. The tilde implies the restriction to single occupancy.
The results are presented in Tables \ref{table3},\ref{table4},\ref{table5}
(for comparison, we also give the results for the Nagaoka state).
We remark from the results for 2 holes that the Fermi distribution
is very close to that of the Nagaoka state (ferromagnetic, $S$=max but in the
$S^z=0$ subspace) with the `Fermi
surface' at approximately $2k_F$ (as in the Nagaoka state) in
distinction with results for finite $J$ where the Fermi vector seems to be at
$k_F$ \cite{hor}.

We can summarize the results from the tables above: i) decreasing $S^z$ the
resemblance of the hole-hole correlations to those of hard core bosons
increases;
for low hole densities a small number of spins flips from the Nagaoka state
suffices to attain the maximum overlap $\simeq 0.99$,
while for larger densities more spin flips are necessary, ii)
the hole pair correlations functions for higher hole densities are closer to
those of hard core bosons than for low densities, iii) the Fermi distribution
function $n_{\sigma}(k)$ for low hole densities is close to that of spinless
fermions (Nagaoka state), iv) decreasing $S^z$ the energy is lowered,
remaining though closer to that of spinless fermions (the $S^z$=max case).

At this point we can discuss our variational argument in light of the
numerical results presented:
our study of two holes corresponds to the low-density limit, where the
`Nagaoka' effect dominates.  The correlations indicate that a fermionic
interpretation is natural and our variational argument is unhelpful.
When we move to higher hole-densities, viz around 33\% , however, the
hard-core boson description becomes much more relevant yielding a
rather accurate description for the charge motion.  The difficulty is
in understanding why the description is so successful, when a
comparison of the energies is so bad.

Usually in variational calculations, the energies are fairly easy to
obtain whereas the wavefunctions are very difficult.  The reason for
this is simply that errors in the wavefunction lead to {\it
quadratic} errors in the energy.  We do not perform a variational
calculation in a technical sense, but are simply using the idea of
the spin-system as playing the role of variational parameters in an
{\it interpretation} of our problem.  Perhaps the simplest way of
explaining this idea is to think about the Schr\"odinger equation for
our chosen representation:

\begin{equation}
E\alpha_c{\vec \Sigma}_c=\sum_{\{c',c\}} \alpha_{c'}
<c'|{\hat T}|c> {\vec \Sigma}_{c'}
\end{equation}

This result is quite general and of little practical analytical value,
although we do use it to perform our numerical calculations.  Our
analytic interpretation comes from the fact that we can `integrate out'
the spin degrees of freedom and obtain an effective `Hamiltonian' for
the charge motion in isolation:

\begin{equation}
E\alpha_c=\sum_{\{c',c\}} \alpha_{c'}
<c'|{\hat T}|c> {\vec \Sigma}_{c'}\cdot {\vec \Sigma}_c
\end{equation}

In concept, this equation is quite similar to that for the hard-core
boson problem.  The particles hop to nearest neighbour sites with a
particular matrix element.  For the hard-core Bose gas, the matrix
element is always $-t$, independent of the configuration of the
bosons.  For the t-model, the matrix element is
$<c'|{\hat T}|c> {\vec \Sigma}_{c'}\cdot {\vec \Sigma}_c$, which ranges
theoretically between $+t$ and $-t$.  For our representation this
matrix element depends on the particular configuration of {\it all the
other particles.}  If this matrix element wildely fluctuated in sign,
then our description would be of no practical help, but in practice,
although this matrix element is reduced from unity, it does not vary
much in magnitude.  This fact successfully explains our variational
results:  The eigenvalue for our system becomes proportional to this
matrix element, while the wavefunction remains almost identical to that
for hard-core bosons.

It also becomes clear what we mean by using the spin system as
`variational parameters'.  The overlaps between the different spin
wavefunctions, ${\vec \Sigma}_{c'}\cdot {\vec \Sigma}_c$, are these
variational parameters in practice.  They must be chosen so as to
keep the phase as uniform as possible, in order to avoid quantum
mechanical phase cancellation, and then further to optimise the
magnitude of
the resulting hopping, subject to the constraint that each spin
configuration is overlapped with many others, each with different
local requirements.

For the eight site model with two holes these overlaps are exactly
uniform (up to the sign necessary to cancel the sign of $<c'|{\hat T}|c>)$),
taking the value ${1\over {\surd 2}}$, but
simultaneously leading exactly to the hard-core bose ground state.
For the larger systems this compensation is not exact, but is an
understandable explanation for our results.

A large amount of work has been devoted to the question of the stability of
the Nagaoka state. There are two kind of approaches to this question:
one is the study of small clusters, the other the calculation of the energy
gain or loss when one spin is flipped from the Nagaoka state. Recent
studies \cite{by,tr,vl} essentially conclude that the Nagaoka state is stable
against a single spin flip.
They also point out that the study of small clusters is rather hazardous
as most often most hole densities correspond to open shells and therefore
it is difficult to obtain good systematics.
Opposite conclusion is reached by high temperature expansion \cite{put}.
We would think however that the case of closed shells might not be
representative
as there is an energy gap appearing due to the finite size of the clusters
which is not representative of the macroscopic system. This way electron-hole
or collective excitations are supressed, thus enhancing the stability of the
Nagaoka state. Second, seen from the variational point of view
above it seems that there is a large variational freedom due to spin flips
and in general the case of a finite density of spin flips should be studied
in order to decide about the stability of the Nagaoka state. Of course the
variational argument above does not exclude the possibility that the
magnetization is finite (even $\sim 1$) in the thermodynamic limit;
if the Nagaoka state turns out to
be stable we should explain why the variational argument does not work.

We can conclude that perhaps the most obvious fact from our small clusters is
that the energies and quantum numbers of small system ground-states are
erratic and dominated by the
particular boundary conditions for the cluster.  It is however possible
to suggest a few `trends' which the calculations exhibit.  Firstly, the
two-hole systems exhibit strong fermionic character.  This is evidenced
by the strong correspondence between the single-(viz n$_\sigma $(k))
correlations for the t-model and spinless fermion model.  All would be
straightforward, if the total-spin of our two-hole ground-states were
large, but it is not.  In fact, as our variational argument would
suggest, the total-spin is zero.  This fact is quite tricky to
interpret, and requires further understanding.  When the
concentration of holes is increased we find rather different behaviour.
Although the particular ground-state spin is erratic, it becomes true
that as the total-spin of the lowest-energy state is reduced, the
overlap with the hard-core Bose ground-state increases.  We can interpret
these results in terms of the competition between the low-concentration
Nagaoka effect and the higher-concentration `bosonization'.  For high
spin the particles gain their energy from single-particle motion as
fermions, while for low spin they gain their energy from collective
motion as hard-core bosons with a reduced probability of hopping
controlled by the spin wavefunction.  At low concentration of holes
the single-particle motion appears more important while at high
concentration the collective motion seems more important.

\acknowledgments
We would particularly like to thank P. Prelovsek for many useful discussions
as well as H. Castella, D. Baeriswyl, J.M.F. Gunn; also the
hospitality of I.S.I. in Torino where this work was started.
This work was supported by the NSF/British Council grant no. 83BC-033384 and
(X.Z) also acknowledges financial support by the
Swiss National Science Foundation under Grant No.
20-30272.90 and the University of Fribourg.

\newpage

\begin{table}
\caption{"Overlaps" (on the left-) and pair correlations functions (on the
right hand side respectively) for N=16 sites}
\begin{tabular}{ccccccccc}
 $N_h$   &   $S^z$       &$\vec k$&   E      &(t$\mid$h.c.b.)&~~~~~~~~~~&
${\vec r}$  & $g(\vec r)_t$  & $g(\vec r)_{h.c.b}$ \\
\tableline
  2    &7&$(\pi/2,0)$&   -6.      &  0.9098&&(1,0)& 0.0040     & 0.0053    \\
       &6&(0,0)      &   -6.2781  &  0.9969&&(1,1)& 0.0078     & 0.0085    \\
       &5&$(\pi/2,0)$&   -6.4117  &  0.9858&&(2,0)& 0.0078     & 0.0085    \\
       &4&(0,0)      &   -6.5238  &  0.9967&&(2,1)& 0.0117     & 0.0103    \\
       &3&$(\pi/2,0)$&   -6.5687  &  0.9892&&(2,2)& 0.0155     & 0.0115    \\
       &2&(0,0)      &   -6.6179  &  0.9966      \\
       &1&$(\pi/2,0)$&   -6.6430  &  0.9908      \\
       &0&(0,0)      &   -6.6775  &  0.9962      \\
 h.c.b &           &(0,0)&   -7.5696  \\
\tableline
  6    &5&$(\pi/2,\pi/2)$&-12.    &0.8955   &&(1,0)& 0.0997     & 0.1012 \\
       &4&$(\pi/2,\pi/2)$&-12.3781 &  0.9595&&(1,1)& 0.1326     & 0.1320 \\
       &3&$(\pi,\pi/2)$  &-12.7418 &  0.9854&&(2,0)& 0.1326     & 0.1320  \\
       &2&(0,0)    &-13.2019 &  0.9932&&(2,1)& 0.1352     & 0.1353     \\
       &1&$(\pi/2,0)$&-13.3512 &  0.9955&&(2,2)& 0.1400     & 0.1359     \\
       &0&(0,0)      &-13.7555 &  0.9980   \\
 h.c.b &           &(0,0)&-16.7881  &           \\
\tableline
  8    &4&$(\pi/2,\pi/2)$&-12.    & 0.8611  &&(1,0)& 0.2034     & 0.2044 \\
       &3&(0,0)       & -13.1416 &  0.9432&&(1,1)& 0.2477     & 0.2433    \\
       &2&$(\pi/2,0)$  &-13.7099 &  0.9787 &&(2,0)& 0.2468     & 0.2433    \\
       &1&(0,0)       &-14.2233 &  0.9874 &&(2,1)& 0.2402     & 0.2441    \\
       &0&(0,0)       &-14.3475  &  0.9919&&(2,2)& 0.2410     & 0.2456    \\
 h.c.b &           &(0,0)&-17.9996    \\
\tableline
\end{tabular}
\label{table1}
\end{table}

\begin{table}
\caption{"Overlaps" (on the left-) and pair correlations functions (on the
right hand side respectively) for N=18 sites}
\begin{tabular}{ccccccccc}
 $N_h$   &   $S^z$       &$\vec k$&   E &  (t$\mid$h.c.b.)&~~~~~~~~~~
& ${\vec r}$  &$g(\vec r)_t$  &$g(\vec r)_{h.c.b}$\\
\tableline
  2    &8&$(2\pi/3,2\pi/3)$&-6.     &0.9212  &&(1,0)& 0.0030     & 0.0042    \\
       &7&$(2\pi/3,0)$  &-6.3160  & 0.9941&&(1,1)& 0.0046     & 0.0064    \\
       &6&(0,0)     &-6.4597  & 0.9946&&(2,0)& 0.0098     & 0.0075    \\
       &5&$(\pi,\pi)$   &-6.4872  & 0.9932&&(2,1)& 0.0077     & 0.0076    \\
       &4&$(2\pi/3,2\pi/3)$&-6.5095  & 0.9915&&(3,0)& 0.0125  & 0.0083    \\
       &3&$(\pi,\pi)$       &-6.5451  & 0.9935       \\
       &2&(0,0)       &-6.5727  & 0.9927       \\
       &1&$(\pi/3,\pi/3)$       &-6.5902  & 0.9922       \\
       &0&(0,0)       &-6.6133  & 0.9920       \\
 h.c.b &           &(0,0)&-7.6212  &              \\
\tableline
  4    &7&$(2\pi/3,2\pi/3)$&-10.    &0.9133&&(1,0)& 0.0258     & 0.0278    \\
       &6&(0,0)       &-10.7118 &  0.9714&&(1,1)& 0.0398     & 0.0402    \\
       &5&$(2\pi/3,0)$&-10.7232 &  0.9901&&(2,0)& 0.0435     & 0.0430    \\
       &4&(0,0)       &-10.8527 &  0.9984&&(2,1)& 0.0461     & 0.0444    \\
       &3&$(\pi,\pi)$ &-10.9915 &  0.9972&&(3,0)& 0.0460     & 0.0450    \\
       &2&(0,0)       &-11.1321  & 0.9980             \\
       &1&(0,0)       &-11.1468  & 0.9967             \\
       &0&(0,0)       &-11.1742 &  0.9979       \\
 h.c.b &           &(0,0)&-13.6111  &                \\
\tableline
\end{tabular}
\label{table2}
\end{table}

\begin{table}
\caption{$n(k)$ for $N=16, N_h=2$, (on the right, for Nagaoka state) }
\begin{tabular}{ccccccccc}
$\pi$  & 0.512 & 0.340 &   0.019 &~~~~~~
$\pi$  & 0.500 & 0.375 &   0.000     \\
$\pi/2$& 0.509 & 0.513 &   0.340 &~~~~~~
$\pi/2$& 0.500 & 0.500 &   0.375     \\
0    & 0.513 & 0.509 &   0.512 &~~~~~~
0    & 0.500 & 0.500 &   0.500     \\
$k_y/k_x$         & 0 & $\pi/2$ &  $\pi$ &~~~~~~
$k_y/k_x$         & 0 & $\pi/2$ &  $\pi$    \\
\tableline
\end{tabular}
\label{table3}
\end{table}

\begin{table}
\caption{$n(k)$ for $N=16, N_h=6$, (on the right, for Nagaoka state) }
\begin{tabular}{ccccccccc}
$\pi$  & 0.212 & 0.096 &   0.066 &~~~~~~
$\pi$  & 0.416 & 0.000 &   0.000     \\
$\pi/2$& 0.650 & 0.212 &   0.096 &~~~~~~
$\pi/2$& 0.500 & 0.416 &   0.000     \\
0    & 0.677 & 0.650 &   0.212 &~~~~~~
0    & 0.500 & 0.500 &   0.416     \\
$k_y/k_x$  & 0   & $\pi/2$ &  $\pi$&~~~~~~
$k_y/k_x$  & 0   & $\pi/2$ &  $\pi$    \\
\tableline
\end{tabular}
\label{table4}
\end{table}

\begin{table}
\caption{$n(k)$ for $N=18, N_h=2$, (on the right, for Nagaoka state) }
\begin{tabular}{ccccccccccc}
$\pi$    &         & 0.497 &              & 0.039 &~~~~~~
$\pi$    &         & 0.500 &              & 0.000   \\
$2\pi/3$ & 0.505 &           &   0.344 &          &~~~~~~
$2\pi/3$ & 0.500 &           &   0.375 &         \\
$\pi/3$  &         & 0.518   &              & 0.497 &~~~~~~
$\pi/3$  &         & 0.500   &              & 0.500 \\
0      & 0.507   &           &   0.505    &  &~~~~~~
0      & 0.500   &           &   0.500    &          \\
$k_y/k_x$& 0   & $\pi/3$ &  $2\pi/3$  &$\pi$  &~~~~~~
$k_y/k_x$& 0   & $\pi/3$ &  $2\pi/3$  &$\pi$  \\
\tableline
\end{tabular}
\label{table5}
\end{table}
\end{document}